\documentclass[english,pra,showpacs,twocolumn]{revtex4-1}
\usepackage[T1]{fontenc}
\usepackage[latin9]{inputenc}
\usepackage{color}
\usepackage{amsmath}
\usepackage{amssymb}
\usepackage{graphicx}

\makeatletter
\usepackage[colorlinks = true,
            linkcolor =blue,
            urlcolor  =blue,
            citecolor = blue,
            anchorcolor = blue]{hyperref}

\makeatother

\usepackage{babel}
\begin{document}
\title{Amplitude-squared squeezing of Schrödinger cat states via Postselected
von Neumann Measurement }
\author{Nuerbiya Aishan, Taximaiti Yusufu}
\author{Yusuf Turek}
\email{yusufu1984@hotmail.com}

\affiliation{School of Physics and Electronic Engineering, Xinjiang Normal University,
Urumqi, Xinjiang 830054, China}
\date{\today}
\begin{abstract}
We know that the original Schrödinger cat states has no amplitude-squared
squeezing. In this paper, we investigate the amplitude-squared squeezing
of Schrödinger cat states using postselected von Neumann measurement.
The results show that after postselected von Neumann measurement,
the amplitude-squared squeezing of Schrödinger cat states change dramatically
and this can be considered a result of weak value amplification. The
squeezing effect also investigated by studying the Wigner function
of Schrödinger cat states after postselected measurement. 
\end{abstract}
\pacs{42.50.-p, 03.65.-w, 03.65.Ta}
\maketitle

\section{\label{sec:1}Introduction}

The squeezing effect plays an essential role in the framework of quantum
theory and its applications. The squeezed states of radiation fields
have reduced uncertainty in specific field quadrature, i.e. quadrature
fluctuations are below the level associated with the vacuum state
or with coherent state \citep{1}. Thus, the squeezed states of radiation
fields which possesses the squeezing effect are considered truly quantum\citep{2}
and have no classical counterpart \citep{3,4}. The study of squeezing,
especially quadrature squeezing of radiation fields has potential
applcaition in optical communication and information theory \citep{10,11,12,13,14,15,16,17,18,7,8,9},
gravitatioanl wave detection \citep{19}, quantum teleportation \citep{19,20,21,22,23,24,25,26,27},
dense coding \citep{28}, resonance fluorscence \citep{29}, and quantum
cryptography \citep{30}. 

With the rapid development of techniques for making higher-order correlation
measurements in quantum optics and laser physics, it is possible to
define the \textcolor{black}{higher-order} squeezing effect of the
field. By considering higher-order correlation functions of the amplitude,
Hong and Mandel \citep{31} defined a state to be squeezed to the
$2Nth$ order if the expectation value of the $2N$th power of the
difference between a field quadrature component and its average value
is less than it would be in a coherent state. Hilley \citep{32,33}
defined another type of higher-order squeezing, named amplitude-squared
squeezing (ASS) of the electromagnetic field. This type of squeezing
arises in a natural way in second-harmonic generation and in a number
of nonlinear optical processes. After that the ASS and more higher-order
squeezing of radiation fields have been investigated in various physical
systems \citep{34,35,36,37,38,39,40,41,42,43,44,45,46,47,48,49,50},
and these theoretical studies have suggested a possibility to extract
information from an optical signal by higher-order correlation measurements.
This has attracted our interest as a means of exploring states which
possess higher-order squeezing. This purpose is related to the generation
and optimization of various quantum states of light; since many of
those states may possess important nonclassical properties like squeezing
and sub-Poissonaian photon statistics. In recent years, significant
attention has been given to this purpose and Schrödinger cat states
are a typical example. Schrödinger cat states which are defined as
the quantum superposition of two coherent states $\vert\alpha\rangle$
and $\vert-\alpha\rangle$ have numerous theoretical and practical
applications in research fields of quantum optics, quantum computation
\citep{51,52}, and quantum information science \citep{53,54,55,56,57}.
Research has shown that the even Schrödinger cat state ($\vert\alpha\rangle+\vert-\alpha\rangle$)
exhibits normal squeezing but not sub-Poissonian statistics, while
the odd Schrödinger cat state ($\vert\alpha\rangle-\vert-\alpha\rangle$)
exhibits sub-Poissonian statistics but has no normal squeezing effect
\citep{58}. Its a well know fact that Schrödinger cat states have
no ASS effect and seem unsuited to related higher-order correlation
measurements. Thus, the question may arise: Is there any method to
amplify or change the inherent properties of a quantum state such
a Schrödinger cat state, so that it can repossess higher-order squeezing?
This question can be addressed through the application of the weak
value amplification technique. 

The weak value amplification technique is related to the postselected
weak measurement proposed by Aharonov, Albert, and Vaidman in 1988
\citep{59}. One of the distinguished properties of weak measurement
compared with traditional projective quantum measurement is that the
induced weak value of the system observable can take large anonymous
values \citep{Aharonov2005}. The feature of the weak value is usually
referred to as an amplification effect of postselected weak measurement,
and can be used to amplify tiny but useful information on physical
systems \citep{61,62,63,64,65,66,67,68,69} . For details of applications
of weak measurement in signal amplification processes, we refer the
reader to the recent overview of the field \citep{70,71}.\textcolor{magenta}{{}
}In weak measurement theory the interaction strength between the system
and the measurement is too weak, and it is enough only to consider
the evolution of the unitary operator up to its first order. However,
if we want to connect the weak and strong measurement, investigate
the measurement feedback of postselected weak measurement, and analyze
experimental results obtained in nonideal measurements, the full-order
evolution of the unitary operator is needed \citep{72,73,74}. This
kind of measurement is referred to as a postselected von Neumann measurement.\textcolor{magenta}{{}
}The postselected von Neumann measurement can be used in state optimization
and precision measurement problems \citep{75,76,77,78,79,80,81}.\textcolor{magenta}{{}
}Recently, one of the authors of this paper investigated the effects
of postselected von Neumann measurement on the properties of single-mode
radiation fields \citep{80} and found that postselected von Neumann
measurement changed the photon statistics and quadrature squeezing
of radiation fields for different anomalous weak values and coupling
strengths.\textcolor{magenta}{{} }However, to the best of our knowledge,
the effects of postselected von Neumann measurement on higher-order
squeezing phenomena of radiation fields have not been previously investigated.

In this work, motivated by our investigations \citep{78,80,81}, we
study the effects of postselected von Neumann measurement on ASS of
Schrödinger cat states.To achieve our goal, we take the spatial and
polarization degrees of freedom of Schrödinger cat states as the measuring
device (pointer) and system, respectively. Along with the standard
process of weak measurement , we take pre- and post-selection on the
system observable. By considering the full-order evolution of the
unitary evolution operator of the total system, we determine the final
state of the pointer after the postselected measurement. After checking
the criteria for existence of ASS of the radiation field, we found
that after using postselected measurement, the ASS of Schrödinger
cat states changed more dramatically than the original one. We plot
the related figures with allowed parameters and the analytical results
indicated that the ASS effects on Schrödinger cat states are caused
by signal amplification of the weak measurement technique. In order
to provide more details of the squeezing phenomena of Schrödinger
cat states after the postselected measurement, we evaluated the Wigner
function for this state.\textcolor{red}{{} }When compared to the initial
state, negative peaks as well as interference structures of Wigner
functions in phase space changed significantly after postselected
measurement.\textcolor{red}{{} }We also found that the shapes of the
Wigner function of Schrödinger cat states are not only squeezed, but
the negative regions increased with increasing coupling strength between
the system and pointer. These results indicated that the nonclassicality
of Schrödinger cat states are more pronounced with large weak values.\textcolor{red}{{}
}We believe that since higher-order correlation measurements are necessary
in some interdisciplinary fields such as quantum biology and quantum
metrology, the current research motivated by the postselected weak
measurement technique is of significant value and may provide some
new, effective methods for implementations of related processes in
those emerging research fields.

The rest of this paper is organized as follows. In Sec. \ref{sec:2},
we outline the first interpretation of amplitude-squared squeezing,
introduce the postselected measurement model and present the normalized
final state of our scheme. In Sec. \ref{sec:3}, we study the amplitude-squared
squeezing of Schrödinger cat states which occur after postselected
von Neumann measurement. In Sec. \ref{sec:4}, we investigate the
Wigner function of Schrödinger cat states after postselected measurement
to explain the squeezing phenomena. Finally, we conclude this work
in Sec. \ref{sec:5}.

\section{\label{sec:2}Fundamental concepts}

\subsection{The definition of squeezing of squared field amplitude}

As early as 1987, M.Hillery showed that squeezing of squared field
amplitude (amplitude-squared squeezing) is a nonclassical effect and
gave some specific examples \citep{33}. Consider a single mode of
electromagnetic field of frequency $\text{\ensuremath{\omega}}$ with
creation and annihilation operator $a^{\dagger},$ $a$. The real
and imaginary parts of the square of the field mode amplitude can
be written as 
\begin{equation}
Y_{1}=\frac{\left(A^{\dagger2}+A^{2}\right)}{2},\ \ Y_{2}=i\frac{\left(A^{\dagger2}-A^{2}\right)}{2},\label{eq:Y12}
\end{equation}
where $A$ and $A^{\dagger}$ are slowly varying operators defined
by $A=e^{i\omega t}\hat{a}$, $A^{\dagger}=e^{-i\omega t}\hat{a}^{\dagger}$,
and obey the same commutation relations as $a$, $a^{\dagger}$. The
operators $Y_{1}$ and $Y_{2}$ satisfy the commutation relationship
\begin{equation}
\left[Y_{1},Y_{2}\right]=i\left(2N+1\right)\text{,}\label{eq:commutation relationship}
\end{equation}
where $N$ is the number operator, $N=A^{\dagger}A$. Thus, the operators
$Y_{1}$ and $Y_{2}$ obey the uncertainty relation
\begin{equation}
\Delta Y_{1}\triangle Y_{2}\ge\langle N+\frac{1}{2}\rangle,\label{eq:uncertanty relation}
\end{equation}
Here, $\triangle Y_{1,2}=\sqrt{\langle Y_{1,2}^{2}\rangle-\langle Y_{1,2}\rangle^{2}}$
denotes the variance of $Y_{1,2}$ under arbitrary state $\vert\phi\rangle$.
We say that the ASS exists in the variable $Y_{i}$ if it satisfies
\begin{equation}
\left(\text{\ensuremath{\triangle}}Y_{i}\right)^{2}\text{\ensuremath{<}}\langle N+\frac{1}{2}\rangle\text{\ \ \ \ for \ensuremath{i=1}or \ensuremath{2}.}\label{eq:squeezing uncertanty}
\end{equation}
In short, the system characterized by the wave function $\vert\phi\rangle$
may exhibit \textcolor{black}{nonclassical} features if it satisfies
Eq. (\ref{eq:squeezing uncertanty}) .

\subsection{Postselected von Neumann measurement and weak value}

In Sec. \ref{sec:1}, we mentioned the applications of postselected
weak measurement. Here, we introduce the main idea of postselected
von Neumann measurement with added related quantities used in our
current work. In quantum measurement theory, the interaction Hamiltonian
can show the main information and relationship between components
to accomplish the measurement process. As the traditional measurement
in the weak measurement case, the coupling interaction between the
system and the measuring device is also given by the standard von
Neumann Hamiltonian \citep{Aharonov2005}

\begin{equation}
H=g\delta(t-t_{0})\hat{\text{\ensuremath{\sigma}}}_{x}\otimes\hat{P},\label{eq:Hamil}
\end{equation}
where $g$ is a coupling constant, $\hat{P}$ denotes the conjugate
momentum operator to the position operator $\hat{X}$ of the measuring
device with $[\hat{X},\hat{P}]=i\hat{I}$, and $\hat{\sigma}_{x}=\vert H\rangle\langle V\vert+\vert V\rangle\langle H\vert$
is an observable of the system that we want to measure. Here, $\vert H\rangle\equiv(1,0)^{T}$
and $\vert V\rangle\equiv(0,1)^{T}$ represent the horizontal and
vertical polarization of the beam, respectively. To guarantee the
precision of a quantum measurement result, the interaction time for
the measuring device and the measured system must be as short as possible.
Thus, for simplicity, we assume the interaction to be impulsive at
time $t=t_{0}$. For this kind of impulsive interaction, the time
evolution operator $e^{-\frac{i}{\hbar}\int Hd\tau}$ of our total
system becomes $e^{-\frac{1}{\hbar}ig\hat{\sigma}_{x}\otimes\hat{P}}$.
Hereafter, we use $\hbar=1$ and assume all factors in $g\hat{\text{\ensuremath{\sigma}}}_{x}\otimes\hat{P}$,
especially $g$, are dimensionless. 

As we know, the weak measurement is characterized by pre- and post-selection
of the system state and a weak value. We assume that initially the
system and measuring device (pointer) are prepared to $\vert\psi_{i}\rangle$
and $\vert\phi\rangle$, and the total initial state can be expressed
as $\vert\psi_{i}\rangle\otimes\vert\phi\rangle$. After the evolution
of the total initial state under the evolution operator $e^{-ig\hat{\sigma}_{x}\otimes\hat{P}}$within
$t_{0}$, we take a postselection with state $\vert\psi_{f}\rangle$
onto $e^{-ig\hat{\sigma}_{x}\otimes\hat{P}}\vert\psi_{i}\rangle\otimes\vert\phi\rangle$,
and obtain the information about a physical quantity $\hat{\sigma}_{x}$
from the final pointer state by the following weak value
\begin{equation}
\langle\hat{\sigma}_{x}\rangle{}_{w}=\frac{\langle\psi_{f}\vert\hat{\sigma}_{x}\vert\psi_{i}\rangle}{\langle\psi_{f}\vert\psi_{i}\rangle},\label{eq:WV}
\end{equation}
This is the definition of the weak value of the system observable.
From Eq. (\ref{eq:WV}), we know that when the preselected state $\vert\psi_{i}\rangle$
and the postselected state $\vert\psi_{f}\rangle$ are almost orthogonal,
the absolute value of the weak value can be arbitrarily large. We
call this feature weak value amplification, and the postselected weak
measurement technique possesses numerous applications as mentioned
in Sec. \ref{sec:1}. 

We can express the position operator $\hat{X}$ and momentum operator
$\hat{P}$, in terms of the annihilation (creation) operators, $\hat{a}$
($\hat{a}^{\dagger}$) in Fock space representation as 
\begin{eqnarray}
\hat{X} & = & \sigma(\hat{a}^{\dagger}+\hat{a}),\label{eq:annix}\\
\hat{P} & = & \frac{i}{2\sigma}(\hat{a}^{\dagger}-\hat{a}),\label{eq:anniy}
\end{eqnarray}
where $\sigma$ is the width of the fundamental Gaussian beam, and
$[\hat{a},\hat{a}^{\dagger}]=\hat{I}$. Thus, we can write the unitary
evolution operator $e^{-ig\hat{\sigma}_{x}\otimes\hat{P}}$ by using
Eq. (\ref{eq:anniy}) as 
\begin{align}
e^{-ig\hat{\sigma}_{x}\otimes\hat{P}} & =\frac{1}{2}(\hat{I}+\hat{\sigma}_{x})\otimes D\left(\frac{\Gamma}{2}\right)+\frac{1}{2}(\hat{I}-\hat{\sigma}_{x})\otimes D\left(-\frac{\Gamma}{2}\right),\label{eq:UNA1}
\end{align}
since the operator $\hat{\sigma}_{x}$ satisfies $\hat{\sigma}_{x}^{2}=\hat{I}$.
Here, the parameter $\Gamma\equiv g/\sigma$, and $D\left(\mu\right)$
is a displacement operator with complex $\mu$ defined by 
\begin{equation}
D(\mu)=e^{\mu\hat{a}^{\dagger}-\mu^{\ast}\hat{a}},\label{eq:DOP}
\end{equation}
Note that $\Gamma$ characterizes the measurement strength, and we
can say that the coupling between the system and pointer is weak (strong)
if $\Gamma<1$$(\Gamma>1)$. We assume through out that the coupling
constant $\Gamma$, is an effective strength of the system and pointer
interaction, and can take all allowed values in weak and strong measurement
regimes. After this von Neumann type postselected measurement, the
final state (not normalized) of the measuring device is given by 
\begin{equation}
\vert\Psi^{\prime}\rangle=\frac{\langle\psi_{f}\vert\psi_{i}\rangle}{2}\left[\left(1+\langle\sigma_{x}\rangle_{w}\right)D\left(\frac{\Gamma}{2}\right)+\!\!\left(1-\!\langle\sigma_{x}\rangle_{w}\right)D\left(\!\!-\frac{\Gamma}{2}\right)\right]\vert\phi\rangle,\label{eq:Fi}
\end{equation}
In the present work, we assume that the pre- and post-selected states
are 
\begin{equation}
\vert\psi_{i}\rangle=\cos\frac{\theta}{2}\vert H\rangle+e^{i\varphi}\sin\frac{\theta}{2}\vert V\rangle,\label{eq:Pre}
\end{equation}
and 
\begin{equation}
\vert\psi_{f}\rangle=\vert H\rangle,\label{eq:Post}
\end{equation}
and then the weak value of the observable $\hat{\sigma}_{x}$ which
defined in Eq. (\ref{eq:WV}) is expressed as 
\begin{equation}
\langle\sigma_{x}\rangle_{w}=e^{i\varphi}\tan\frac{\theta}{2}.
\end{equation}
Here, $\theta\in[0,\pi]$ and $\varphi\in[0,2\pi)$. As mentioned
earlier, the weak value can take an anomalous value and it is accompanied
by low successful postslection probability $P_{s}=\vert\langle\psi_{f}\vert\psi_{i}\rangle\vert^{2}=\cos^{2}\frac{\theta}{2}$. 

\section{\label{sec:3} The ASS of Schrödinger cat states}

In this section we study ASS of Schrödinger cat states after postselected
von Neumann measurment. The Schrödinger cat state is a typical quantum
\textcolor{black}{state which} is composed of the superposition of
two coherent correlated states moving in opposite directions. This
has many applications in quantum information processing. In this study,
we take the spatial and polarization degrees of freedom of Schrödinger
cat states as the measuring device and system, respectively, and investigate
the effects of postselected measurement on the polarization and spatial
components of the Schrödinger cat state beams. The mathematical expression
of normalized Schrödinger cat states are \citep{Agarwal2013} 
\begin{equation}
\vert\Phi\rangle=K\left(\vert\alpha\rangle+e^{i\omega}\vert-\alpha\rangle\right),\label{eq:SCSs}
\end{equation}
 where 
\begin{equation}
K=[2+2e^{-2\vert\alpha\vert^{2}}\cos\omega]^{-\frac{1}{2}}.
\end{equation}
 is the normalization coefficient, and $\alpha=\vert\alpha\vert e^{i\delta}$
is an arbitrary complex number with modulus $\vert\alpha\vert$ and
argument $\delta$. Here, we would like to mention that $\omega\in\left[0,2\pi\right]$,
when $\omega=0\left(\omega=\pi\right)$ it is called the even (odd)
Schrödinger cat state, and when $\omega=\frac{\pi}{2}$ it is also
called the Yurke-Stoler state .

By using the fundamental concepts introduced in Sec. \ref{sec:3}(B),
we can get the final normalized state of the pointer after changing
$\vert\phi\rangle$ in Eq. (\ref{eq:Fi}) to the Schrödinger cat states
$\vert\Phi\rangle$, and this can be expressed as 

\begin{align}
\left|\Psi\right\rangle  & =\frac{\kappa}{2}\left[\left(1+\langle\sigma_{x}\rangle_{w}\right)D\left(\frac{\Gamma}{2}\right)+\left(1-\langle\sigma_{x}\rangle_{w}\right)D\left(-\frac{\Gamma}{2}\right)\right]\left|\Phi\right\rangle ,\label{eq:fi}
\end{align}
 with normalization constant 
\begin{align}
\kappa & =[\frac{1}{2}(1+\vert\langle\sigma_{x}\rangle_{w}\vert^{2})+K^{2}(1-\vert\langle\sigma_{x}\rangle_{w}\vert^{2})cos(2\Gamma Im[\alpha])e^{-\frac{\Gamma^{2}}{2}}\nonumber \\
 & +\frac{K^{2}}{2}\Re[\left(1-\vert\langle\sigma_{x}\rangle_{w}\vert^{2}-2iIm[\langle\sigma_{x}\rangle_{w}]\right)\nonumber \\
 & \left(e^{i\omega}e^{-\frac{1}{2}\vert2\alpha+\Gamma\vert^{2}}+e^{-i\omega}e^{-\frac{1}{2}\vert2\alpha-\Gamma\vert^{2}}\right)]]^{-\frac{1}{2}}.\label{eq:normalization coefficient}
\end{align}
 To investigate the ASS of Schrödinger cat states, we have to follow
the condition mentioned in the previous section about the existence
of the ASS of a single mode electromagnetic field with frequency $\text{\ensuremath{\omega}}$.
After some simple algebra, the condition for existence of ASS of a
single mode radiation field, Eq. (\ref{eq:squeezing uncertanty}),
can be rewritten as 

\begin{align}
R & =\left(\triangle K_{1}\right)^{2}-\langle a^{\dagger}a+\frac{1}{2}\rangle\nonumber \\
 & =\frac{1}{2}Re[\langle a^{4}\rangle]+\frac{1}{2}\langle a^{\dagger2}a^{2}\rangle-\left(Re[\langle a^{2}\rangle]\right)^{2}<0.\label{eq:15}
\end{align}
 Here, $\langle.\rangle$ denotes the expectation values of corresponding
quantities under the state $\vert\Psi\rangle$. It can be seen that
the negativity of the variable $R$ reveals the ASS phenomenon of
the state. In this paper, we only examine the $Y_{1}=\frac{1}{2}\left(a^{\dagger2}+a^{2}\right)$
component of the field. To achieve our goal, first of all we have
to calculate the above related quantities and their explicit expressions
are listed below.

\begin{widetext}

1.The expectation value $\langle a^{2}\rangle$ under the state  $\vert\Psi\rangle$
is given by

\begin{align}
\langle a^{2}\rangle & =\frac{\kappa{}^{2}K^{2}}{4}\left[\vert1+\langle\sigma_{x}\rangle_{w}\vert^{2}H_{1}\left(\Gamma\right)+\vert1-\langle\sigma_{x}\rangle_{w}\vert^{2}H_{1}\left(-\Gamma\right)+\left(1-\langle\sigma_{x}\rangle_{w}\right)\left(1+\langle\sigma_{x}\rangle_{w}\right)^{\ast}H_{2}\left(\Gamma\right)+\left(1+\langle\sigma_{x}\rangle_{w}\right)\left(1-\langle\sigma_{x}\rangle_{w}\right)^{\ast}H_{2}\left(-\Gamma\right)\right],\label{eq:a2}
\end{align}
with 
\begin{align*}
H_{1}\left(\Gamma\right) & =2\left(e^{-2\vert\alpha\vert^{2}}cos\omega+1\right)\left(\alpha^{2}+\frac{\Gamma^{2}}{4}\right)-2ie^{-2\vert\alpha\vert^{2}}\Gamma\alpha sin\omega,
\end{align*}
and 
\[
H_{2}\left(\Gamma\right)=e^{2i\Gamma Im[\alpha]}e^{-\frac{\Gamma^{2}}{2}}\left(\alpha-\frac{\Gamma}{2}\right)^{2}+e^{i\omega}\left(\alpha+\frac{\Gamma}{2}\right)^{2}e^{-2\vert\alpha+\frac{\Gamma}{2}\vert^{2}}+e^{-i\omega}\left(\alpha-\frac{\Gamma}{2}\right)^{2}e^{-2\vert\alpha-\frac{\Gamma}{2}\vert^{2}}+e^{-2i\Gamma Im[\alpha]}\left(\alpha+\frac{\Gamma}{2}\right)^{2}e^{-\frac{\Gamma^{2}}{2}}.
\]
2.The expectation value $\langle a^{\dagger2}a^{2}\rangle$ under
the state $\vert\Psi\rangle$ is given by

\begin{align}
\langle a^{\dagger2}a^{2}\rangle & =\frac{\kappa{}^{2}K^{2}}{4}\left[\vert1+\langle\sigma_{x}\rangle_{w}\vert^{2}K_{1}\left(\Gamma\right)+\vert1-\langle\sigma_{x}\rangle_{w}\vert^{2}K_{1}\left(-\Gamma\right)+2Re\left[\left(1-\langle\sigma_{x}\rangle_{w}\right)\left(1+\langle\sigma_{x}\rangle_{w}\right)^{\ast}\left(K_{2}\left(\Gamma\right)+K_{2}\left(-\Gamma\right)\right)\right]\right],
\end{align}
with 
\begin{align*}
K_{1}\left(\Gamma\right) & =\vert\alpha+\frac{\Gamma}{2}\vert^{4}+\vert\alpha-\frac{\Gamma}{2}\vert^{4}+2Re\left[e^{i\omega}\left(\alpha^{\ast}+\frac{\Gamma}{2}\right)^{2}\left(-\alpha+\frac{\Gamma}{2}\right)^{2}e^{-2\vert\alpha\vert^{2}}\right],
\end{align*}
and 
\[
K_{2}\left(\Gamma\right)=e^{i\omega}\vert\alpha+\frac{\Gamma}{2}\vert^{4}e^{-2\vert\alpha+\frac{\Gamma}{2}\vert^{2}}+e^{2i\Gamma\Im(\alpha)}e^{-\frac{\Gamma^{2}}{2}}\left(\alpha^{\ast}+\frac{\Gamma}{2}\right)^{2}\left(\alpha-\frac{\Gamma}{2}\right)^{2}.
\]
3.The expectation value $\langle a^{4}\rangle$ under state $\vert\Psi\rangle$
is given by
\begin{align}
\langle a^{4}\rangle & =\langle\Psi\vert a^{4}\vert\Psi\rangle\nonumber \\
 & =\frac{\kappa{}^{2}K^{2}}{4}\left[\vert1+\langle\sigma_{x}\rangle_{w}\vert^{2}G_{1}\left(\Gamma\right)+\vert1-\langle\sigma_{x}\rangle_{w}\vert^{2}G_{1}\left(-\Gamma\right)+\left(1-\langle\sigma_{x}\rangle_{w}\right)\left(1+\langle\sigma_{x}\rangle_{w}\right)^{\ast}G_{1}\left(\Gamma\right)+\left(1+\langle\sigma_{x}\rangle_{w}\right)\left(1-\langle\sigma_{x}\rangle_{w}\right)^{\ast}G_{2}\left(-\Gamma\right)\right],
\end{align}
with
\begin{align*}
G_{1}\left(\Gamma\right) & =2\alpha^{4}+3\Gamma^{2}\alpha^{2}+\frac{\Gamma^{4}}{8}+2\cos\omega e^{-2\vert\alpha\vert^{2}}\left(\alpha^{4}-2\Gamma\alpha^{3}+\frac{3}{2}\Gamma^{2}\alpha^{2}-2\frac{\Gamma^{3}}{4}\alpha+\frac{\Gamma^{4}}{16}\right),
\end{align*}
 and 
\begin{align*}
G_{2}\left(\Gamma\right) & =e^{2i\Gamma\Im\left(\alpha\right)}\left(\alpha-\frac{\Gamma}{2}\right)^{2}e^{-\frac{\Gamma^{2}}{2}}+e^{i\omega}\left(\alpha+\frac{\Gamma}{2}\right)^{2}e^{-2\vert\alpha+\frac{\Gamma}{2}\vert^{2}}+e^{-i\omega}\left(\alpha-\frac{\Gamma}{2}\right)^{2}e^{-2\vert\alpha-\frac{\Gamma}{2}\vert^{2}}+e^{-2i\Gamma\Im\left(\alpha\right)}\left(\alpha+\frac{\Gamma}{2}\right)^{2}e^{-\frac{\Gamma^{2}}{2}}.
\end{align*}

\end{widetext}

\begin{figure}
\includegraphics[width=8cm]{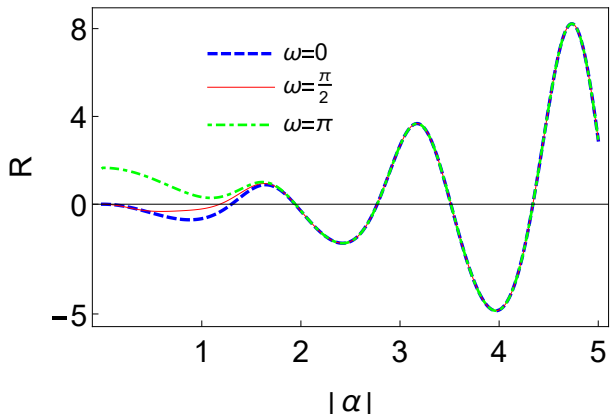}

\caption{\label{fig:1}ASS of Schrödinger cat states. Here, we take $\theta=\frac{\pi}{2}$,
$\delta=0$, $\varphi=\frac{7\pi}{9}$, $\Gamma=2$.}
\end{figure}

To check the effects of postselected von Neumann measurement on ASS
of Schrödinger cat states, we plot $R$ as a function $\vert\alpha\vert$
for different $\omega$ and the analytical results are shown in Fig.
\ref{fig:1}. We know that there is no ASS of initial Schrödinger
cat states. However, as indicated in Fig. \ref{fig:1}, the $R$ can
be below zero after postselected von Neumann measurement and can change
dramatically for large values of $\vert\alpha\vert$ with large weak
values. Fig. \ref{fig:1} also shows that $R$ can take negative values
periodically, and $R$ of three kinds of Schrödinger cat states have
the same trend with increasing $\vert\alpha\vert$. 

\begin{figure}
\includegraphics[width=8cm]{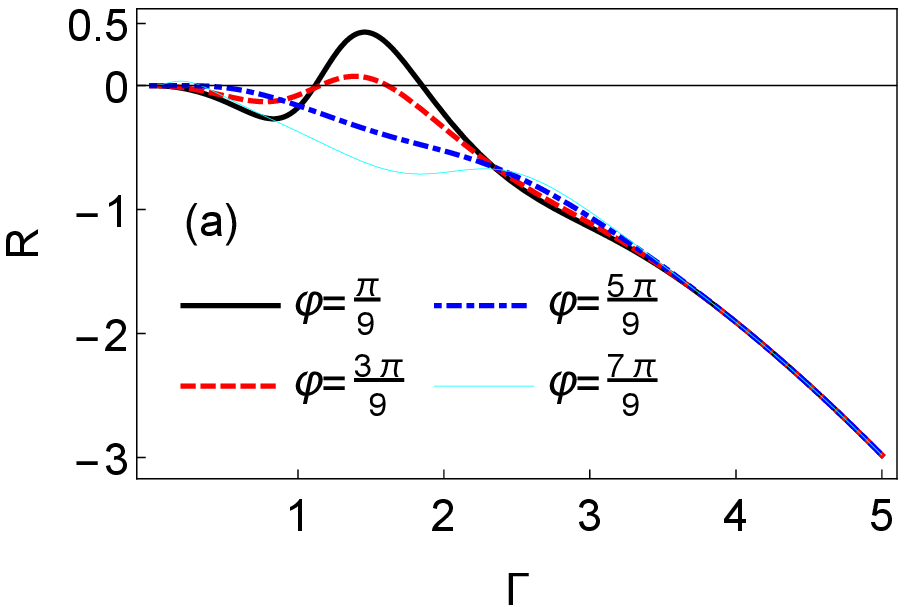}

\includegraphics[width=8cm]{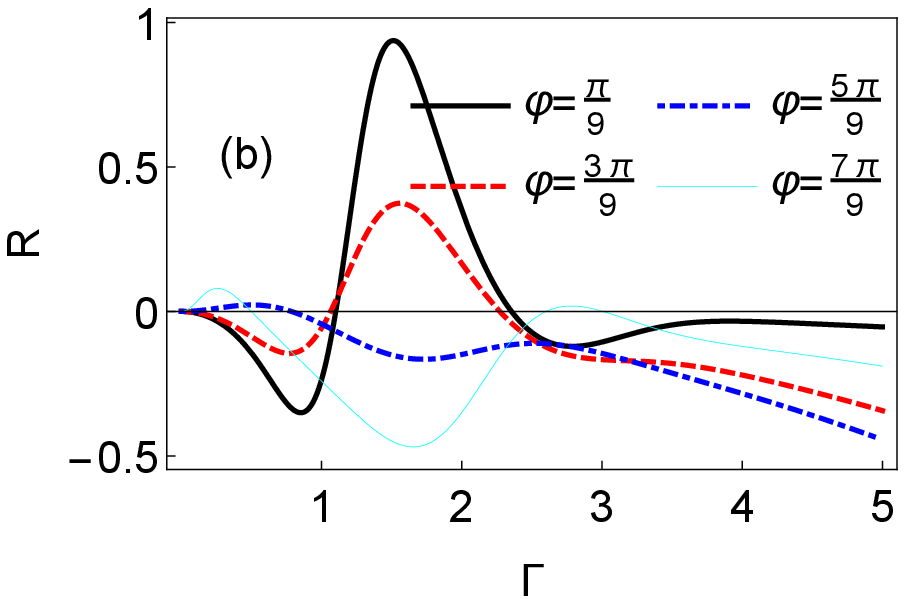}

\includegraphics[width=8cm]{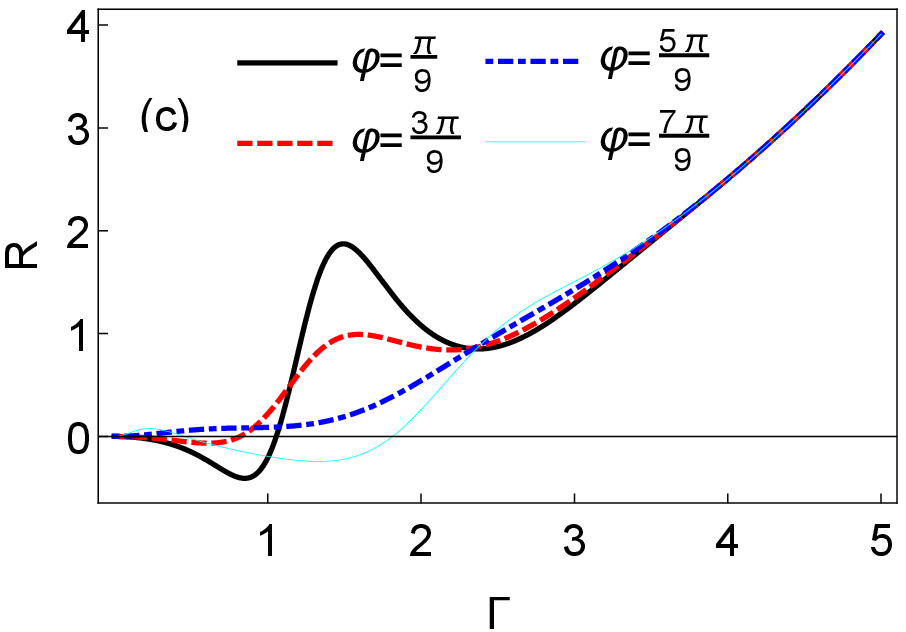}

\caption{\label{fig:2}ASS of Schrödinger cat states. Here, we take $\vert\alpha\vert=1$
and other parameters are the same as those used in Fig. \ref{fig:1}
(a) $\omega=0$; (b) $\omega=\frac{\pi}{2}$; (c) $\omega=\pi$.}
\end{figure}

As mentioned in Sec. \ref{sec:2}, in our scheme the coupling strength
$\Gamma$ can take any values in weak and strong measurement regimes.
To further confirm our claims, we plotted the variation curves of
$R$ as a function of $\Gamma$ for different weak value for the even
(odd) Schrödinger cat state and Yurke-Stoler state, respectively.
It is clearly shown in Fig. \ref{fig:2} that there is no ASS when
the interaction strength $\Gamma=0$ (no interaction), for all three
Schrödinger cat states. This is proof of the fact that there is really
no ASS for Schrödinger cat states initially. From Fig. \ref{fig:2}(a)
we observe that the ASS of the even Schrödinger cat state behaves
more and more strongly as the interaction strength $\Gamma$ increases.
Fig. \ref{fig:2}(b) show the $R$ of the Yurke-Stoler state. We observed
that in most of the regions, $R$ takes on negative values, especially
for large real values, and the curves all trend to below zero as $\Gamma$
increases. This means that the ASS of the Yurke-Stoler state behaves
more stably for strong measurement with large weak values. On the
contrary, we saw in Fig. \ref{fig:2}(c) that although the variable
$R$ takes negative values when $\Gamma$ takes relatively small values,
the ASS of odd Schrödinger cat state totally disappear when $\Gamma$
increases, regardless of the change of the weak value. In addition,
it is evident from Fig. \ref{fig:2} that increasing of the weak value
has a positive effect on the ASS phenomenon of Schrödinger cat states,
which we believe also results from the signal amplification effect
of the weak value. 

In our previous work \citep{80}, we discussed the effects of postselection
von Neumann measurements on the ordinary squeezing properties of Schrödinger
cat states. The results showed that the quadrature squeezing (ordinary
squeezing) effect of the Schrödinger cat states increases with increasing
interaction strength for anomalous weak values compared to the initial
pointer state. Therefore, combined with the above results, we believe
that the dramatic changes brought by the postselected von Neumann
measurements on the odinary and second-order squeezing of the Schrödinger
cat state should not be underestimated.

In order to better explain the effects of the postselected von Neumann
measurement on the nonclassicality of Schrödinger cat states including
ordinary and ASS effects, in the next section we will discuss the
Wigner function of Schrödinger cat states with state $\vert\Psi\rangle$.

\section{\label{sec:4}Wigner Function of Schrodinger cat states after postselected
measurement}

The Wigner function is the earliest quasi-probability distribution
function in phase space, and it is also a very interesting structure
in quantum mechanics. The state when the Wigner function takes negative
values in phase space is a nonclassical state, and it is direct pro
of the nonclassicality of the state. To yield more information on
the squeezing effect of Schrödinger cat states after postselected
measurement, we investigate their Wigner functions in phase space.
In general, Wigner function is defined as the two-dimensional Fourier
transform of the symmetric order characteristic function, and the
Wigner function for the state $\rho=\vert\Psi\rangle\langle\Psi\vert$
corresponding to Eq. (\ref{eq:fi}) can be written as \citep{Agarwal2013}
\begin{equation}
W(z)\equiv\frac{1}{\pi^{2}}\int_{-\infty}^{+\infty}\exp(\lambda^{\ast}z-\lambda z^{\ast})C_{N}(\lambda)e^{-\frac{\lambda^{2}}{2}}d^{2}\lambda,\label{eq:35-1}
\end{equation}
where $C_{N}(\lambda)$ is the normal ordered characteristic function,
and is defined as 
\begin{equation}
C_{N}(\lambda)=Tr\left[\rho e^{\lambda a^{\dagger}}e^{-\lambda a}\right].\label{eq:34}
\end{equation}
 By substituting the final normalized pointer state $\vert\Phi\rangle$
into Eq. (\ref{eq:35-1}), we can give the explicit expression of
the Wigner function as

\begin{align*}
W\left(z\right) & =\frac{\kappa{}^{2}K^{2}}{2\pi}[\vert1+\langle\sigma_{x}\rangle_{w}\vert^{2}F_{1}\left(\Gamma\right)+\vert1-\langle\sigma_{x}\rangle_{w}\vert^{2}F_{1}\left(-\Gamma\right)\\
 & +2\Re[\left(1-\langle\sigma_{x}\rangle_{w}\right)\left(1+\langle\sigma_{x}\rangle_{w}\right)^{\ast}F_{2}\left(\Gamma\right)],
\end{align*}
with 
\begin{align*}
F_{1}\left(\Gamma\right) & =e^{-\frac{1}{2}\Gamma^{2}}\left(e^{-2\vert z+\alpha\vert^{2}}e^{2\Gamma Re[z+\alpha]}+e^{-2\vert z-\alpha\vert^{2}}e^{2\Gamma Re[z-\alpha]}\right)\\
 & +2e^{-2\vert z\vert^{2}}e^{-2\left(\frac{\Gamma^{2}}{4}-\Gamma Re[z]\right)}f\left(\Gamma,\omega\right),
\end{align*}
and 
\begin{align*}
F_{2}\left(\Gamma\right) & =e^{-2\vert z-\alpha\vert^{2}}e^{2i\Gamma Im[z]}+e^{-2\vert z+\alpha\vert^{2}}e^{-2i\Gamma\left(Im[z]+iRe[\alpha]+Im[\alpha]\right)}\\
 & +2e^{-2\vert z\vert^{2}}e^{-2\Gamma\left(Im[\alpha]-Re[\alpha]\right)}f\left(\Gamma,\omega\right).
\end{align*}
Here, $f\left(\Gamma,\omega\right)=cos\left(4Im[z]Re[\alpha]+4Re[z]Im[\alpha]-2\Gamma Im[z]-\omega\right)$,
and $z=x+ip$ is a complex variable in phase space. In general, this
Wigner function is real, and it is bounded $-\frac{2}{\pi}\le W\left(z\right)\le\frac{2}{\pi}$.
If $\Gamma=0$, it is reduced to the Wigner function of initial Schrödinger
cat states which are characterized by the state $\vert\Phi\rangle$,
in Eq. (\ref{eq:SCSs}).

\begin{widetext} 

\begin{figure}
\includegraphics[width=15cm]{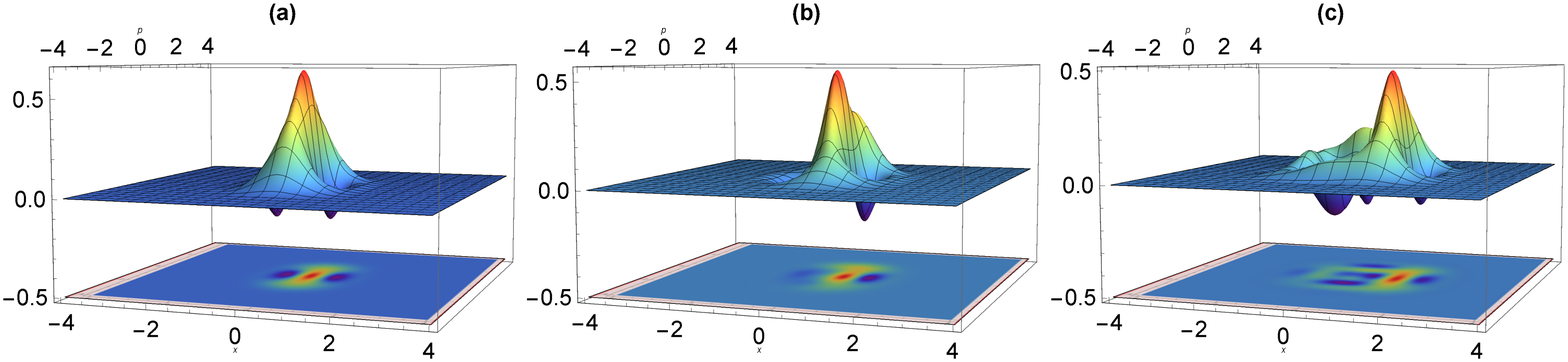}

\includegraphics[width=15cm]{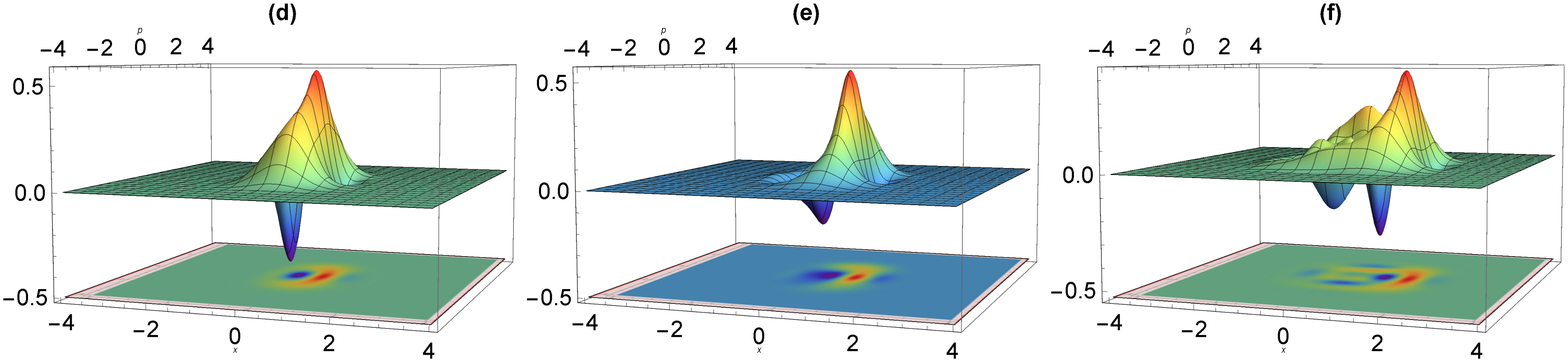}

\includegraphics[width=15cm]{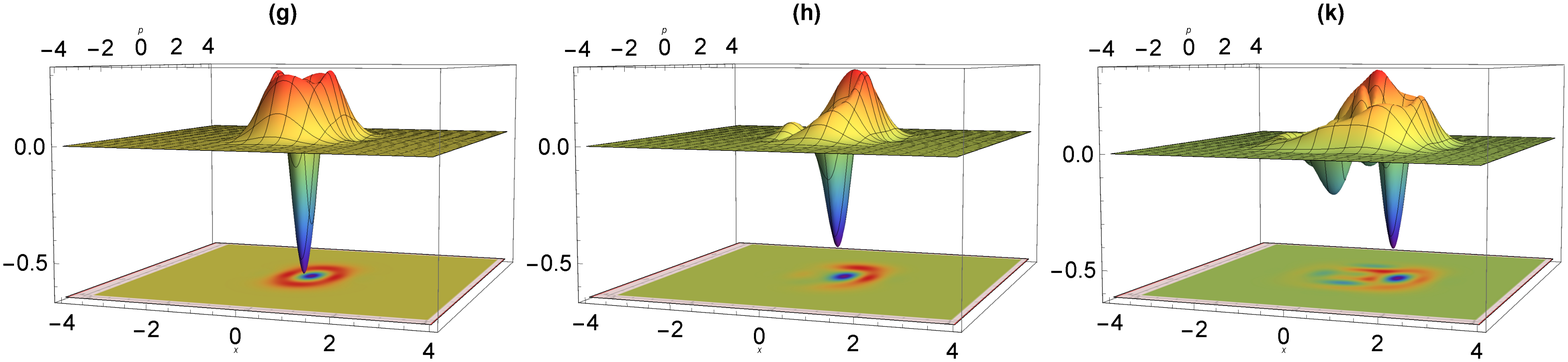}

\caption{\label{fig:3}Wigner function of Schrödinger cat states after postselected
von Neumann measurement with changing parameters. Each column represents
different measurement strengths with $\Gamma=0,0.5,2$, and are ordered
accordingly from left to right. Figures (a) to (c) correspond to the
even Schrödinger cat state ($\omega=0$); (d) to (e) correspond to
the Yurke-Stoler state ($\omega=\frac{\pi}{2}$); and (f) to (k) correspond
to the odd Schrödinger cat state ($\omega=\pi$). Here, we take $\vert\alpha\vert=1$,
$\theta=\frac{\pi}{2}$, $\delta=0$, $\varphi=\frac{7\pi}{9}$.}
\end{figure}

\end{widetext}

We plot in Fig. \ref{fig:3}, the Wigner functions with different
parameters of interest for the Schrödinger cat states as mentioned
above . Each row represents the even (odd) Schrödinger cat state and
the Yurke-Stoler state, respectively. The three plots in each row
from left to right represents the $\Gamma=0,0.5,2$, and we also took
the large weak value ($\text{\ensuremath{\varphi=\frac{7}{9}\pi}}$)
corresponding to the most pronounced ASS phenomenon in Fig. \ref{fig:2}.
The Wigner functions in Fig. \ref{fig:3} all exhibit redundancy and
highly nonclassical characteristics in phase space. By comparing the
three plots in each row, we can see that the shapes of the Wigner
functions are not only squeezed as the measured intensity $\Gamma$
increases, but we can clearly see the quantum interference structures
formed between the peaks. Moreover, by comparing each histogram we
can see that there are some differences in the peaks of the Wigner
functions for different states at the same interaction strength $\Gamma$.
In the strong measurement region $\left(\Gamma=2\right)$, the Wigner
functions of the three types of Schrödinger cat states exhibited the
most pronounced coherence properties.

In addition to this, one can observe the color of the shadow for the
peaks on the lower plane and can judge the sizes of the corresponding
negative values, with the darker color corresponding to a more negative
value of $W\left(z\right)$. The negative regions of the Wigner functions
of the Schrödinger cat states increased in strong measurement regions,
and these results obtained by the Wigner function corroborate our
observations in Sec. \ref{sec:3}. In short, we can achieve a better
ASS effect as the nonclassicality of the state increases.

\section{\label{sec:5}Conclusion and remarks}

We know that the ASS phenomenon does not exist in the initial Schrödinger
cat states. In this paper, we investigated the effects of postselected
von Neumann measurement on the ASS of Schrödinger cat states. For
our purpose, we first give a mathematical expression for the final
state of the measured pointer, and then find the exact expression
for the variable $R$ associated with ASS by calculations and show
the analytical curves for different system parameters. However, through
the analysis, we observed a dramatic change in the ASS of the Schrödinger
cat states after considering the measurement which is characterized
by postselection and weak value. We found that as the interaction
strength $\Gamma$ increases, the ASS phenomenon of three types of
Schrödinger cat states exhibit interesting behavior. Among them, the
ASS of the even Schrödinger cat state behaves more positively after
the measurement, while the odd Schrödinger cat state behaves exactly
the opposite way. Moreover, the larger the weak value is, the more
pronounced the ASS is under the same conditions. This indicates that
the signal amplification effect of the weak values play an important
role in this process.

To explain the squeezing effects of Schrödinger cat states more clearly,
we reconstructed the Wigner functions for each state after postselected
measurement and analyzed the Wigner functions for different related
parameters. It was observed that not only are the peaks squeezed but
they also exhibit coherent structure in specific coupling strength
regions. Among them, the Wigner function of the even Schrödinger cat
state exhibits the most obvious coherence properties after postsletected
measurement. 

In the present work, the results clearly showed that the postselected
von Neumann measurement has a positive effect on the nonclassicality
of the Schrödinger cat states, especially on its ASS. Therefore, we
hope that the results of the present study can provide new methods
to study the quantum information processes related to ASS of radiation
fields.

In this study, we only investigate the effects of postselected measurements
on the second-order squeezing of Schrödinger cat states, and it is
belong to the state optimization process. State optimization is a
effective method to increase the implementation efficiency of related
processes. Thus, in order to expand the practical applications of
postselected von Neumann measurements, it would be interesting to
study the effects of postselection von Nuemman measurement on the
properties of a wide variety of quantum states including entanglement,
noise reduction and state control. Work along these lines are currently
in progress, and we anticipate that our results will be presented
in the near future.
\begin{acknowledgments}
This work was supported by the Natural Science Foundation of Xinjiang
Uyghur Autonomous Region (Grant No. 2020D01A72), the National Natural
Science Foundation of China (Grant No. 11865017) and the Introduction
Program of High Level Talents of Xinjiang Ministry of Science.

\end{acknowledgments}

\bibliographystyle{apsrev4-1}
\addcontentsline{toc}{section}{\refname}\bibliography{ref}

\end{document}